\documentclass[aps,pra,reprint,groupedaddress,showkeys]{revtex4-1}

\usepackage{graphicx}
\usepackage{amssymb,amsmath}
\usepackage{subfig}
\graphicspath{ {images} }

\usepackage[dvipsnames]{xcolor}
\usepackage{csquotes}
\usepackage{float}
\newcounter{inlineenum}
\renewcommand{\theinlineenum}{\roman{inlineenum}}
\newenvironment{inlineenum}
{\unskip\ignorespaces\setcounter{inlineenum}{0}%
	\renewcommand{\item}{\refstepcounter{inlineenum}{\textit{\theinlineenum})~}}}
{\ignorespacesafterend}

\begin{document}

\title{Moving efficiently through a crowd: a nature inspired traffic rule}



\author{Danny Raj M}
\email[]{dannym@iisc.ac.in}
\affiliation{Department of chemical engineering, Indian Institute of Science Bangalore, India}

\author{Kumaran V}
\email[]{kumaran@iisc.ac.in}
\affiliation{Department of chemical engineering, Indian Institute of Science Bangalore, India}


\date{\today}

\begin{abstract}

In this article, we propose a traffic rule inspired from nature, that facilitates an elite agent to move efficiently through a crowd of inert agents. When an object swims in a fluid medium or an intruder is forced through granular matter, characteristic flow-fields are created around them. We show that if inert agents, made small movements based on a traffic rule derived from these characteristic flow-fields, they efficiently reorganize and transport enough space for the elite to pass through. The traffic rule used in the article is a dipole-field which satisfactorily captures the features of the flow-fields around a moving intruder. We study the effectiveness of this dipole traffic rule using numerical simulations in a 2D periodic domain, where one self-propelled elite agent tries to move through a crowd of inert agents that prefer to stay in a state of rest. Simulations are carried out for a wide range of strengths of the traffic rule and packing density of the crowd. We characterize and analyze four regions in the parameter space---free-flow, motion due to cooperation, frozen and collective drift regions---and discuss the consequence of the traffic rule in light of the collective behavior observed. We believe that the proposed method can be of use in a swarm of robots working in constrained environments.

\end{abstract}


\maketitle

\section{Introduction}
\label{intro}
As population increases, so does the average number of vehicles on roads. This leads to large commuting times, traffic jams, unsafe driving conditions etc. that makes traveling on road quite frustrating. In densely populated countries, where the throughput of traffic is high, lane-flow is often not enforced: in order to use up the extra space available on the roads. This is particularly the case in most roads in India where the traffic is dense and heterogeneous (consisting of two, three, four wheeler and sometimes even long trucks). In these dense-traffic flows, it is not uncommon for emergency vehicles like ambulances, rushing to a hospital with a patient, or a fire-rescue mobile speeding towards an accident zone, to be slowed down considerably or even get stuck in the traffic.
Delays in reaching the spot, could mean life or death in cases depending on the severity of the need of the emergency service. Roadway-infrastructures that are in place in developing countries are not ambulance friendly, resulting in a bottle-neck for the health services\cite{TOI2014, DH2016, HT2016, TNIE2017}.

Nevertheless, a skilled ambulance driver tries to maneuver through the crowd, while the vehicles around the ambulance try to make small movements to make enough space for the ambulance to move through. From the perspective of traffic management, one might ask: how should the vehicles coordinate and move to maximize the mobility of the ambulance? A common solution that drivers resort to under such circumstances is to create an empty lane in front of the ambulance, wide enough to allow it to pass through. This is done with the vehicles moving to the sides or away from the ambulance. However, this is possible only for the case where the density of the traffic is small enough for the vehicles to make space for an extra lane. When densities are high, ambulances often remain stuck, waiting for the traffic to be released at a nearby junction\cite{Sundar2015} (a behaviour which is also observed in experiments with pedestrians who anticipate and respond to an intruder\cite{Nicolas2019}).
How then should vehicles move under dense traffic conditions? Is there a traffic rule that the vehicles can follow to make small coordinated movements such that the re-arrangement creates enough space for the ambulance to pass through? Here, a traffic rule, refers to the movement strategy that a vehicle can follow; typically a function of the state of the surrounding traffic and the state of the vehicle with priority.


\paragraph*{Motion in crowded environments:}
In search of a traffic rule that maximizes the mobility of the ambulance (elite agent), we turn to the several known natural and artificial systems where we find motion in crowded environments. Swimmers move through a fluid by displacing the fluid around them \cite{VanHouwelingen2019}; the size/mass of the swimmer (even a micro-swimmer) is several orders of magnitude greater than the molecules of the fluid that are displaced\cite{Spagnolie2012}. An example with similar features, where the sizes are comparable, can be found in the field of granular physics\cite{Candelier2010, Kolb2014}. A solid intruder is driven at a constant force or velocity, through a granular medium that is either at rest or under steady vibration. Grains are displaced locally around the intruder and the resultant dynamics leads to fluctuations in the velocity of the intruder or the force required to maintain the same velocity. This constitutes the field of active micro-rheology where the local structure of the medium is characterized from the measured fluctuations. In addition, when the intruder used, has a structural asymmetry like in a tapered rod (fore-aft), the intruder motion in a vibrated granular bed is spontaneous due to the difference in rate of collisions between the front and the rear\cite{Kumar2014}. Motion of these rods are mediated via consistent re-arrangement of grains around the rod as it passes through. When many rods are used, the grain re-arrangement around each of the rods can result in long range interactions between other rods, that can give rise to flocking.

Intruder motion through granular medium results in strong re-arrangements of the inert grains around the intruder. With increasing density, intruder-motion undergoes a fluidization transition from smooth to intermittent motion before finally exhibiting jamming. Another interesting example of motion through a crowded environment is the experiments of Tennenbaum et al \cite{Tennenbaum2016}, where they study the rheology of a suspension of fire ants. These ants under low stress can interlock with its neighbors to exhibit a solid like behavior and then open-up under the application of a high loading to behave more fluid-like, exhibiting a non-Newtonian behavior. When a heavy object is placed on top of a suspension of army ants packed in a column, they start to open up their inter-lock and facilitate the motion of the heavy object through the dense suspension. One would notice that the motion of ants are in such a way that they move away from the front of the object while continuously filling up its wake. These ants exhibit motion to allow the passage of the heavy object.

\paragraph*{Grains and traffic}: 
There are many differences between grains and vehicles that one should have in mind before attempting to transfer ideas from granular physics to traffic dynamics. Vehicles have a heading direction: movement is along the direction in which the vehicle has oriented or its reverse; vehicles turn to change its direction. Grains, on the other hand, do not have such a heading and they flow or move in the direction where the force acts. 
Also, typical traffic scenarios exhibit non-reciprocity in the inter-agent force: when B approaches A from behind, only B is expected to slow down without any change in the behavior in A\cite{Helbing2004}. That is, the forces between two nearby agents A and B are not equal and opposite $ (\mathbf{F}_{A-B} \neq \mathbf{F}_{B-A}) $. Since, grains do not have this heading direction which is necessary to define what the front and back of an agent is, we do not consider this feature in the model presented in this article.

The solution to the problem of moving through a crowd is non-obvious even for the case where the agents move like grains. In addition to being a simpler starting point, it bears relevance to a system where the agents are capable of movement in all directions, operated in a constrained environment (like drones). Hence, in the remainder of the article, we focus our attention to agents that do not have an orientation and investigate the problem of identifying and evaluating the traffic rule that an inert crowd of agents can follow to facilitate large movement of an elite agent.

\section{Model}
\subsection{Deriving a nature-inspired traffic rule} \label{NatInsTraffRule}
In granular physics (and other complex flows too), one can characterize the micro-structure of the granular medium and its response to local stress, by observing the motion of an intruder through the medium \cite{Squires2005a, Khair2006, Wilson2009, Wilson2011, Reichhardt2015, Zia2018}. By posing a problem where we search for a traffic rule that maximizes the mobility of the intruder or elite (objective), we are in fact solving an inverse problem where our interest is in determining the \textit{optimal microrheology} of the crowd. An intruder driven through a granular medium generates a characteristic flow of grains around it, which can be visualized by averaging the particle velocities over time. It is a flow-field with a \textit{source-sink} like feature due to the emptying of grains from the frontal region of the intruder while simultaneously filling its rear end (fig 6b in ref \cite{Candelier2010}). A simple functional expression that can capture the trend qualitatively is a \textit{dipole field}, let's call this $U$. Note that the velocity fields observed in real granular flow scenarios do not necessarily have a $ 1/r^2 $ dependence or a fore-aft symmetric decay. 

\begin{figure}
	\centering
	\includegraphics[width=1\linewidth]{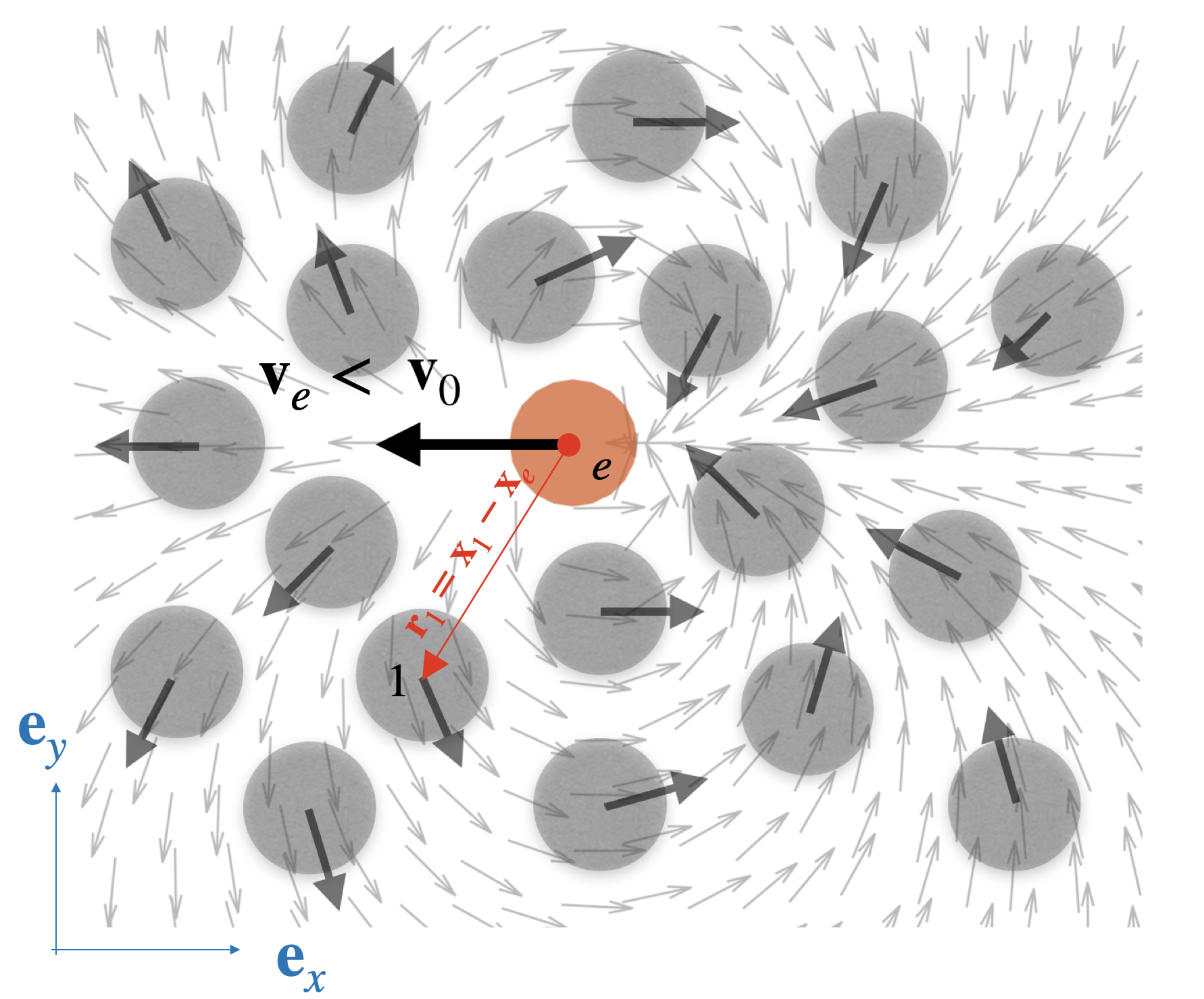}
	\caption{Dipole field ($ U(\mathbf{r}_j) $ as in Eq \ref{dipolefield}), plotted around the elite agent and the directions of the force and the resultant motion of the inert agents according to the dipole traffic rule, when $\mathbf{v}_e<\mathbf{v}_0$. The arrows here represent the direction and not the magnitude (which scales as $ \frac{1}{r^2} $, where $ r $ is the distance of the inert from the elite). The vector displacement w.r.t. the elite, $\mathbf{r}_j$, is marked for a single inert agent.}
	\label{fig:dipoletrafficrule}
\end{figure}

\paragraph*{Traffic rule:} One can define a traffic rule based on this dipole field: \begin{inlineenum}
    \item imagine this fictitious field $U$ to be around the elite agent (e),
    \item inert agents (j) respond to this field through a force on them proportional to $U$ evaluated at the position of the inert.
\end{inlineenum}
$U$ is given as in Eq \ref{dipolefield} (corresponding to two-dimensional flows) where, $ \mathbf{v}_0 $ is the desired velocity of the elite agent whose instantaneous velocity is $ \mathbf{v}_e $, $ \mathbf{r}_j $ is the position vector of the inert with respect to the elite and $ r = \| \mathbf{r}_j\|$ (as illustrated in figure \ref{fig:dipoletrafficrule}).

\begin{eqnarray}
    \mathbf{U}(\mathbf{r}_j)=\frac{(\mathbf{v}_e-\mathbf{v}_0)}{r^2}.\Bigg[\boldsymbol{\delta_{ij}}-2 \frac{\mathbf{\mathbf{r}}_j \otimes \mathbf{\mathbf{r}}_j}{r^2}\Bigg]
    \label{dipolefield} \\
    \text{where, } \mathbf{r}_j=\mathbf{x}_j- \mathbf{x}_e \nonumber
\end{eqnarray}

Note that when the elite agent travels at the desired velocity, the strength of the dipole field drops to zero. This way, the motion of the inert is induced by the traffic rule only when the elite is unable to move in its desired velocity, thereby minimizing the need for the motion of inert whenever possible. 
From the perspective of an optimization formulation, this choice of implementation of the dipole traffic rule, implicitly encodes for two objectives simultaneously: maximising the mobility of the elite agent at minimal or no movement of the inert agents.

\subsection{Particle based model}
Agents in traffic flows can be modeled in a way similar to grains in granular flows. A discrete particle approach to granular matter, involves modeling grains as particles with finite size, defined by the positions and velocities of their center of masses. Their positions are updated based on the instantaneous velocities, which evolve based on Newton's second law of motion. Hence to solve this problem, a complete knowledge of the different forces on the grain is necessary: external, inter-particle, particle-boundary, etc. All the forces experienced by grains have a physical origin that include friction, inertia, gravity, incompressibility etc. While agents in traffic flow exhibit some of these characteristics like inertia, impenetrability etc., their motion is primarily behavioral in nature---where agent tries to move with a velocity it desires or avoids collision with neighbors etc. This behavioral maneuvering can be explained with what are called, \textit{social forces}\cite{Helbing1995}.

We model agents as finite sized, circular particles (with radius $R$) moving on a 2D plane under the influence of social forces. Eq \ref{kinematic-sfm} shows how the position of an agent $ \mathbf{x}_j $ is updated based on the knowledge of its velocity $ \mathbf{v}_j $. Eq \ref{sfm-full} is the Newton's second law with the social forces for the agents. Agents experience three different types of forces in our model. 
\begin{inlineenum} 
	\item \textit{restitution force} that always attempts to bring the agent to its desired state of rest or motion
	\item \textit{collision-avoidance force} that prevents collision between agents and the 
	\item \textit{dipole force} inspired from nature, which induces or coordinates the motion of inert around the elite.
\end{inlineenum}

\begin{equation}
\frac{d\mathbf{x}_j}{dt}=\mathbf{v}_j \label{kinematic-sfm}
\end{equation}

\begin{subequations} \label{sfm-full}
	\begin{equation}\label{force-inert-sfm}
	m_i\frac{d\mathbf{v}_i}{dt}=-m_i\frac{\mathbf{v}_i }{\tau}+ \sum_{\forall j\neq i} \mathbf{F}_{ij}^r + K \mathbf{U}_i
	\end{equation}
	
	\begin{equation}\label{force-elite-sfm}
	m_e\frac{d\mathbf{v}_e}{dt}=m_e\frac{\mathbf{v}_0-\mathbf{v}_e}{\tau}+ \sum_{\forall j\neq e} \mathbf{F}_{ej}^r
	\end{equation}
	
	\begin{gather}\label{resistance-sfm}
	\mathbf{F}_{ij}^r = 
	\begin{cases}
	-\gamma_r (\mathbf{d}_{ij}-2R)^{-(B+1)}\hat{\mathbf{d}}_{ij} & |\mathbf{d}_{ij}|<l_{cr}, \mathbf{\dot{d}}_{ij}>0 \\
	0 & \mbox{otherwise}
	\end{cases}\\
	\text{where, } \mathbf{d}_{ij} = \mathbf{x}_i-\mathbf{x}_j \text{ and } \mathbf{\dot{d}}_{ij}=\mathbf{v}_i \cdot \mathbf{d}_{ji} + \mathbf{v}_j \cdot \mathbf{d}_{ij}\nonumber
	\label{inter-agent}
	\end{gather}
\end{subequations}

The first terms in the R.H.S. of Eq \ref{force-inert-sfm} and \ref{force-elite-sfm} , are the \textit{restitution} social force for the inert and elite respectively, similar to that used in ref\cite{Helbing2000, Helbing2000a}. In the absence of all other forces, the inert agents will stop moving while an elite agent will move with a velocity $\mathbf{v}_0$, in $ \mathcal{O}(\tau) $ seconds. The motivation of the agents to move in certain directions with a preferred speed, is what makes the system of agents active; energy is pumped into the system through this term.
The second term is the inter-agent interaction force, that brings in the agent's reaction to a nearby neighbor, to avoid collision. The third term (only seen in Eq \ref{force-inert-sfm})), incorporates the effect of the traffic rule explained in section \ref{NatInsTraffRule}. The dipole field is conceptualized as a force, $ \mathbf{F}_{dipole} = K \mathbf{U} $, where $ K $ is the strength of the dipole force ($ K=0 $ represents the case with no traffic rule). The motion of the elite agent is modeled in a similar fashion, as shown in Eq \ref{force-elite-sfm}. The only differences are the desired velocity $ \mathbf{v}_0 $ instead of the state of rest and the absence of the traffic rule. In the swarm robotics community, the approach to modeling collective behavior discussed here, is termed as \textit{physicomimetics} \cite{Spears2004}.

Agents avoid collisions with other agents by responding to their approach. $ \mathbf{F}_{ij}^r $ characterizes the force between agents $ i \text{ and } j $ when they are in close proximity and approaching each other. The nature of the force as defined in Eq \ref{resistance-sfm} does not allow the circular agents (with area $ 2\pi R^2 $) to overlap: force $ \to \infty $ when $ |\mathbf{d}_{ij}|\to 2R $. The dependence of the magnitude of force on the distance between agents $ |\mathbf{d}_{ij}| $ is defined by the parameter $ B $. If the value of $ B $ is very high, the interaction between agents tends to become hard-sphere, like in a granular medium. However, in traffic flow scenarios, one would expect the agents to respond to approaching agents in advance and slow down to prevent a collision (due to the behavioral origin of the social force). To model such a behavior, one has to choose a smaller value for $ B $, but care should be taken to eliminate long range correlations that may arise. Hence, we employ $ B=2 $ in our simulations (similar to \cite{Helbing2000a}), and use a cut-off length $ l_{cr} = 3R$, to make an agent interact only with its immediate neighbors. 

In addition to proximity, we consider this force only when agents approach each other, \textit{i.e.} $ \mathbf{\dot{{d}}}>0 $. As a result, agents moving with the same velocity will not experience a force even if they are in close proximity, allowing them to have a heterogeneous density field in a crowded system. If the interaction had been purely repulsive and not dependent on whether or not the agents approached each other, then the interaction would have attempted to smooth-out the density variations in the system. This feature in the interaction is common in real dense traffic, where agents respond to neighbors only when they approach. An agent decelerates slowly when the approaching agents are far away, a faster response when they are in close proximity and no response when not approaching. When the agents are crowded, in a slowly drifting jammed state, they do not try to move away from each other, but only move in desired direction while avoiding collisions. 

\section{Results and Discussion}

\begin{figure}
	\centering
	\includegraphics[width=1.0\linewidth]{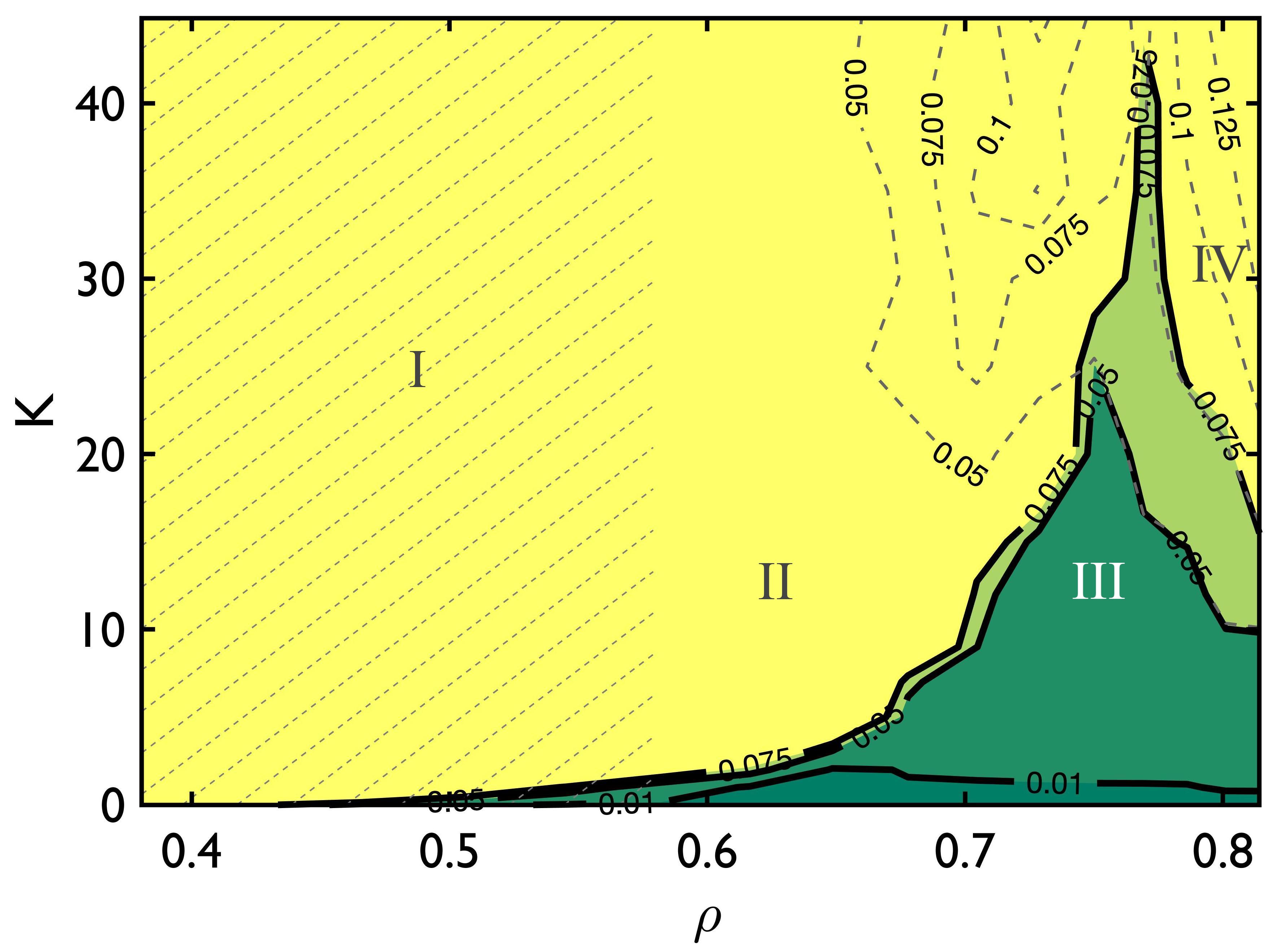}
	\caption{The density--dipole-field strength ($K-\rho$) parameter space can be divided into four regions: I -- free flow, II -- motion with traffic rule, III -- frozen and IV -- collective drift. Thick lines -- total mobility and thin lines -- only due to collective drift.}
	\label{fig:paramaterspaceregularpacking}
\end{figure}

To test the effectiveness of the dipole traffic rule, simulations are carried out for a range of packing densities $ \rho\in (0.2, 0.85) $ and dipole strengths $ K\in[0,50] $, in a 2D periodic domain (unit cell width and length, $ W\times L $), consisting of $150$ agents which include $1$ elite agent (other system sizes are also attempted to make sure that there is no loss of generality). Two types of initial configurations of agents are used in our simulations; a regular triangular and a disordered random packing (obtained from \cite{Specht2018}). Unless specified, the results shown in the article are for the case of regular initial arrangements. The initial configurations are subject to fluctuating forces with collision avoidance to perturb the agents from any symmetric arrangements. The resultant assembly of agents then serves as the initial condition for the simulations with the traffic rule. 
The desired direction and speed of the elite agent are set as $ -x $ direction and unity respectively, as shown in Eq \ref{des_dir_elite}. 
\begin{equation}
    \mathbf{v}_0=\mathbf{e}_{-x}
    \label{des_dir_elite}
\end{equation}
The parameters, relaxation time scale $\tau$ and the mass $ m_i\text{ and }m_e $ of the agents, are chosen to be unity without loss of generality. The dynamics of the traffic problem depends on the parameter $\gamma_r$ which is the strength of the inter-agent interaction. However, the qualitative features of the results presented in the article are expected to remain unchanged.

Based on the dynamics of traffic flow, we identify four regions in the parameter space, highlighted in figure \ref{fig:paramaterspaceregularpacking} (see ESI for video). 
\begin{enumerate}
    \item A \textit{free-flow} region corresponding to low packing densities, where motion of the elite is possible even in the absence of the dipole force (hashed region, marked as I), 
    \item \textit{Motion under cooperation} region where the dipole traffic rule is essential for movement of the elite (region II), 
    \item \textit{Frozen} region where motion of the elite is insignificant (thick solid lines mark the different contours of mobility, in region III) and 
    \item \textit{Collective drift} region (IV) where the local frozen micro structure determines the flow direction and speed of drift of the entire assembly (thin solid lines mark the mobility due to drifting, which is found in some parts of region II as well).
\end{enumerate}

In what follows, we analyze the movement of the elite in these four different regions in the parameter space. We compute the mobility of the elite, and probe into the movement of the inert agents in light of the collective phenomena exhibited.

	

\subsection{Free flow regime}

When packing densities are less than approximately $ 0.5$, the elite agent forces its way through the inert crowd even in the absence of the dipole traffic rule. As the elite travels through the crowd, it encounters inert agents that behave as obstacles to the motion of the elite, thereby reducing its mobility $ \mathcal{M} $, defined as the time-averaged velocity of the elite agent in the principle direction of motion (see, Eq \ref{Mobility_X}). 

\begin{equation}
	\mathcal{M} =  \frac{\int_{0}^{T}{\mathbf{v}_e \cdot \mathbf{e}_{-x} dt}}{T} \label{Mobility_X}\\
\end{equation}

The mobility $ \mathcal{M} $ quantifies the net-movement in the direction of $\mathbf{v}_0$, averaged over time.
In addition to slowing down, interaction with the inert results in drifting $ \mathcal{D} $ along the direction perpendicular to the principle direction of motion. We define $\mathcal{D}$ as the average absolute velocity in that direction as shown in Eq \ref{Drift_Y_freeflow}.

\begin{equation}
	\mathcal{D} =\frac{\int_{0}^{T}{|\mathbf{v}_e \cdot \mathbf{e}_y| dt}}{T} \label{Drift_Y_freeflow}
\end{equation}

The drift $ \mathcal{D} $ quantifies the average excursions of the elite in a direction transverse to that of the intended one ($\mathbf{e}_{\pm y}$). When the elite agent travels in the direction of $\mathbf{v}_0$, at all time, $\mathcal{D}$ takes a value of $0$ and $ \mathcal{M} $ takes the value of $1$.

\begin{figure}
    \centering
    \includegraphics[width=1\linewidth]{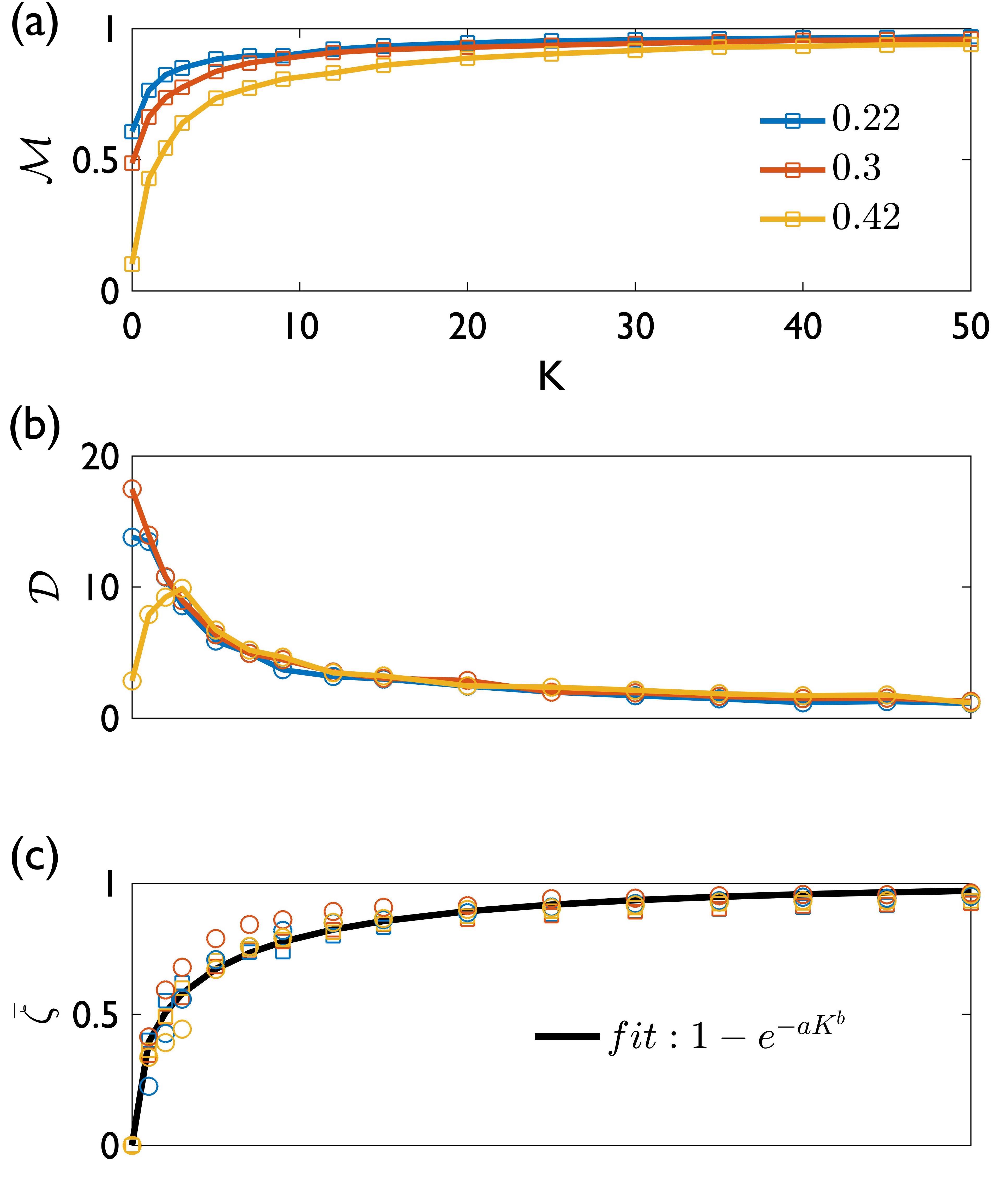}
    \caption{Mobility characteristics in the free-flow region. (a) - Mobility $\mathcal{M}$ as a function of $K$ for different values of the density; (b) - Drift $\mathcal{D}$ as a function of $K$ for different values of the density; (c) - Normalized drift and mobility $\bar{\zeta}$ as a function of K for different values of density, collapses onto a single curve}
    \label{fig:RegionI-mobility}
\end{figure}

In the presence of the traffic rule, the inert agents move around the elite, resulting in a reduction in the interactions between the elite and the inert agents and thereby reducing the resistance. As a result, with increasing $ K $, mobility increases (with $ \mathcal{M}|_{K\to \infty}\to 1$, see figure \ref{fig:RegionI-mobility} (a)). 
However, drifting exhibits a non-monotonic behaviour with $K$. We find that $\mathcal{D}$ initially increases and then it decreases with $K$ as $ \mathcal{D}|_{K\to \infty}\to 0$ (see figure \ref{fig:RegionI-mobility}(b)).
This is because, in the absence of any traffic rule $(K=0)$, motion of the elite in the principle direction is strongly reduced due to head-on collisions with the neighbouring inert agents, which subsequently reduces the total movement in the  perpendicular direction. However, the ratio of drift to mobility ($\mathcal{D}/\mathcal{M}$), is found to be monotonically decreasing.
We observe the normalized mobility and drift-mobility ratio, $\bar{\mathcal{M}}$ and $\bar{(\frac{\mathcal{D}}{\mathcal{M}})}$ computed as in Eq \ref{normalizing_freeflow}, to be invariant to packing density, depending only on $ K $ as, $ \big[ 1-\exp\big({-a K^{b}\big)} \big]$ (where $ a,b=0.5 $) in the free-flow regime.
\begin{eqnarray}
    \label{normalizing_freeflow}
	\bar{\zeta}(K)=\frac{\zeta(\rho,K) - \zeta(\rho,0)}{{\zeta(\rho,\infty)} - \zeta(\rho,0)} \\
	\text{where, } \zeta=\frac{\mathcal{D}}{\mathcal{M}} \text{ or } \mathcal{M} \text{ and } \rho \lesssim 0.5 \nonumber
\end{eqnarray}

\begin{figure*}
	\centering
	\includegraphics[width=0.85\linewidth]{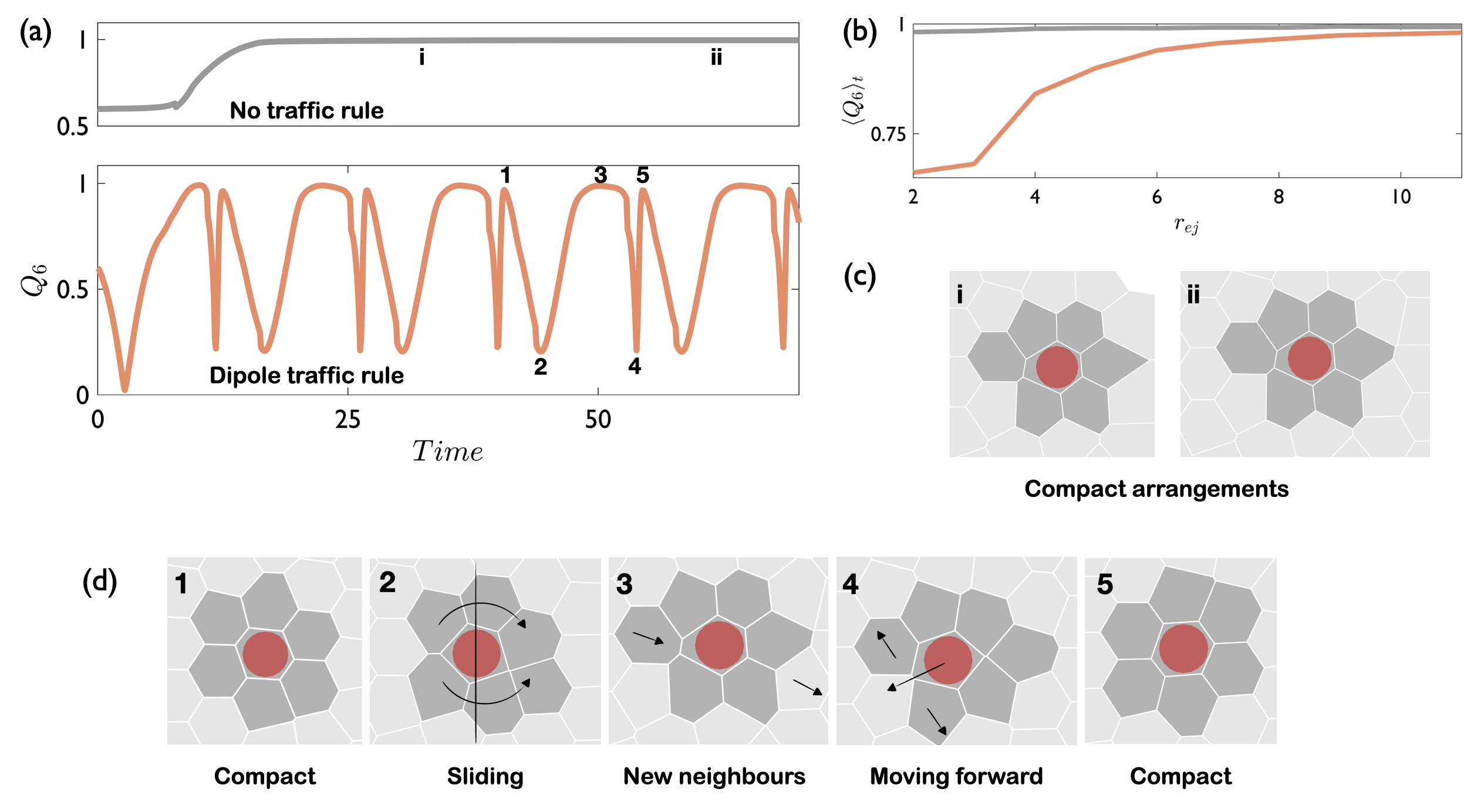}
	\caption{Crowd-characteristics at high densities, where motion is possible only with the dipole traffic rule. (a)- $ q_6 $ order parameter for the neighbors around the elite agent as a function of time for $ K={0,12} $ of $\rho= 0.73$. We observe fluctuations in the order parameter when the traffic rule is in place; (b)- Time-averaged $ q_6 $ order parameter as a function of distance from the elite $ (r_{ej}) $ for $ K={0,12} $ of $\rho= 0.73$, shows melting of the assembly close to the elite; (c)- Local crowd configuration around the elite for the case without any traffic rule; (d)- Dynamically evolving local crowd configuration around the elite as the dipole traffic rule coordinates the re-arrangement of the inert agents to facilitate the motion of the elite.}
	\label{fig:q6_timeseries_timeavg}
\end{figure*}

\subsection{Frozen vs motion with traffic rule}
As the density of packing is increased $ (>0.5) $, the motion of the elite agent is strongly constrained and the system eventually becomes \textit{frozen} (which corresponds to $ \mathcal{M} \leq 0.075 $). When completely frozen and in the absence of the traffic rule, the entire system consisting of both the elite and the inert, drifts with a velocity $ \mathbf{v}_i=\mathbf{v}_0/N $, and with a corresponding $\mathcal{M}=1/N$ due to the forcing of the elite agent. The arrangement of the agents around the elite becomes compact and triangular.
One could evaluate the $ q_6 $ order parameter, defined as $\big\langle \sum_{k \in \eta_j} \exp(-6\theta_{jk} \rm{i}) \big\rangle_j$, where $ \eta_j $ consists of the $6$ nearest neighbours of the agent $j$.
When the arrangement around an agent is triangular (close packed), the order parameter takes the value of unity. When the assembly is frozen, the corresponding $ q_6 $ averaged over all the agents, becomes $\approx 1$. Presence of the dipole traffic rule, does not guarantee motion of the elite as the system may still be frozen. As the strength $ K $ of the dipole force is increased, the inert agents around the elite begin to make enough movements, to give rise to the motion of the elite.

If one were to compute the $ q_6 $ order parameter for the inert agents, just around the elite agent, it is quite evident that the crystallized agents \textit{melt}, to make the motion of the elite agent possible as seen in figure \ref{fig:q6_timeseries_timeavg}a.
We observe periodic oscillations in the $q_6$ parameter as the elite moves through the inert crowd, when the dipole rule is turned on. In contrast, when the traffic rule is not in place, the $q_6$ parameter rises to a value close to unity and stays unchanged. 
We also observe that the melting is local. A time-averaged $ q_6 $ order parameter, plotted as a function of distance from the elite $ r_{ej} $ (see, figure \ref{fig:q6_timeseries_timeavg}b) shows, a reduced $ (\langle q_6 \rangle_t) $ near the elite which increases monotonically to higher values far away from the elite in the presence of the traffic rule. This suggests that the change in configuration of the agents that facilitates motion of the elite is primarily local in nature.

To understand how the changes in local configuration facilitates the motion of the elite, we examine the position and motion of the immediate neighbours of the elite agent at various time points, as the elite cruises through the inert crowd. 
Figure \ref{fig:q6_timeseries_timeavg}c, shows the compact triangular arrangement of the inert around the elite when the assembly is frozen. Without the traffic rule, the elite tries to move forward and very soon results in tight arrangements that restricts any further motion (i and ii in figure \ref{fig:q6_timeseries_timeavg}a, c).
On the other hand (see figure \ref{fig:q6_timeseries_timeavg}d), when the dipole traffic rule is switched on, the agents initially form a compact configuration which arises as a result of the forward motion of the elite through the crowd. Once this motion is no longer possible, due to the local compaction of the assembly, the inert agents begin to slide around the elite (due to the traffic rule). 
This gives rise to more of the inert agents in the rear of the elite than its front, breaking the fore-aft symmetry and resulting in a smaller value of $q_6$ parameter (see configurations 1 to 3 in figure \ref{fig:q6_timeseries_timeavg}a, d).
As the inert moves around the elite, it opens up space in front of the elite and introduces new neighbours for the elite to interact with changing the immediate configuration around it. 
Once sufficient space is created, the elite agent quickly moves to the new configuration (4-5 transition in figure \ref{fig:q6_timeseries_timeavg}a, d) to result in another compacted local arrangement and the process continues.
One should note that several other re-arrangement events as the elite moves, and periods of drifting can make the configuration change much more complicated. However, the series of steps described in figure \ref{fig:q6_timeseries_timeavg}d conveys the major events that occur during the motion of elite in the presence of a traffic rule at high densities.

\begin{figure*}
    \centering
    \includegraphics[width=1\linewidth]{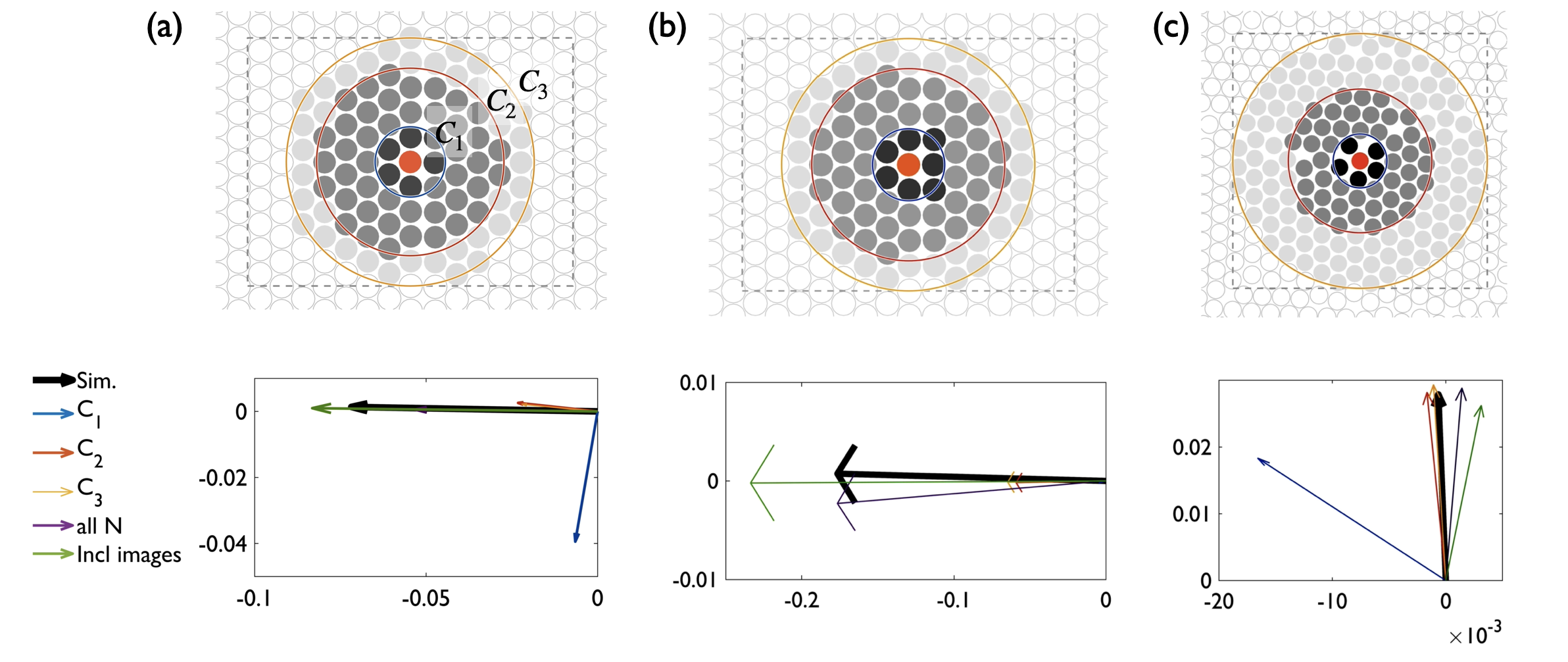}
	\caption{Micro-structure of the frozen assembly (top panels) and the corresponding drift direction (lower panels) as observed in the numerical simulations and that which is calculated based on Eq \ref{coll-drift-sum-final} using varying amounts of information of the micro-structure (agents within $C_1, C_2, C_3$, all $N$ agents and including their images).}
	\label{fig:Colldriftall}
\end{figure*}

\subsection{Collective drift}\label{sec:colldrift}
At very high packing densities, when the assembly of agents is frozen, the only motion expected is a slow drift, where the elite agent pushes the all the agents with a velocity $ \| \mathbf{v}_k \| = \frac{\|\mathbf{v}_0\|}{N} $; this is indeed observed when $ K $ is small. However, interestingly at high values of $ K $, we observe a more significant drift (in the direction $ \mathbf{e}_{-x} $, for regularly packed initial configurations) where the $ \| \mathbf{v}_k \| \gg \frac{\|\mathbf{v}_0\|}{N} $. If the drifting observed is due to the dipole traffic rule, then the features of drifting, namely the magnitude and direction of the drift velocity, should be a function of the local packing, or the frozen microstructure. Here, the microstructure of the frozen assembly refers to the relative displacement of the inert agents around the elite: $\mathbf{r}_j$ (as defined in Eq \ref{dipolefield}).
Simulations with a randomly packed assembly of agents (initial configuration), reveals that the direction can be arbitrary and dependent on the arrangement of the inert agents around the elite. When the density of packing is such that the system exhibits intermittent dynamics, we observe drifting to occur during the periods when the elite is caged and every time the cage breaks and the elite moves to a new one, there is major restructuring of the inert agents around the elite which results in the emergence of a new drift direction every time (see ESI for video).

To understand the relation between the steady state drift velocity $ \mathbf{v}_s $ and the microstructure of the frozen assembly, we can go back to the governing equations Eq \ref{force-inert-sfm} and \ref{force-elite-sfm}. Assuming steady state and frozen-\enquote*{ness} of the assembly---which means that the the direction of the drift is a constant and there is no change in the microstructure---one can derive a relation between $ \mathbf{v}_s $ and $ \mathbf{I}_{ij} $. 
\begin{equation}
    \mathbf{I}_{ij} = \boldsymbol{\delta_{ij}}-2 \frac{\mathbf{\mathbf{r}}_j \otimes \mathbf{\mathbf{r}}_j}{r^2}
    \label{microstructure}
\end{equation}
Here the tensor $ \mathbf{I}_{ij} $, is purely a function of the frozen microstructure $\mathbf{r}_j$, and quantifies the contributions of the local arrangement on the dipole traffic rule (Eq \ref{microstructure} is same as the term within the square brackets in Eq \ref{dipolefield}).
When the assembly is frozen, the structure of the assembly remains unchanged. Hence, we can neglect the force between agents (collision avoidance) and sum all the forces up as shown in Eq \ref{coll-drift-sum}, where $ \mathbf{v}_i $ becomes the drift velocity $ \mathbf{v}_s $ (note: in our study we have assumed $m_e = m_i = 1$).
Substituting Eq \ref{dipolefield} in \ref{coll-drift-sum}, gives rise to Eq \ref{coll-drift-sum-sl1}. With further simplification one can arrive at Eq \ref{coll-drift-sum-final} which shows the relation between the frozen crowd-microstructure and the drift velocity.

\begin{subequations}
	\begin{equation}\label{coll-drift-sum}
	\sum_{\forall j\neq e} \Big(K \mathbf{U}_j -\frac{\mathbf{v}_s}{\tau}\Big) + \frac{\mathbf{v}_0-\mathbf{v}_s}{\tau} = 0
	\end{equation}
	
	\begin{equation}\label{coll-drift-sum-sl1}
	\sum_{\forall j\neq e} \Big(K(\mathbf{v}_s - \mathbf{v}_0) \mathbf{I}_{ij} -\frac{\mathbf{v}_s}{\tau}\Big) + \frac{\mathbf{v}_0-\mathbf{v}_s}{\tau} = 0
	\end{equation}
	
	\begin{equation}\label{coll-drift-sum-final}
	\mathbf{v}_s=\mathbf{v}_0 \cdot \Big(K\tau \sum_{\forall j\neq e}  \mathbf{I}_{ij} - \delta_{ij}\Big) \Big(K\tau \sum_{\forall j\neq e} \mathbf{I}_{ij} - N\delta_{ij}\Big)^{-1}
	\end{equation}
\end{subequations}

When the strength of the traffic rule is very high, $ K\to \infty $, $ K\tau \sum_{\forall j\neq e}  \mathbf{I}_{ij} \gg N\delta_{ij} > \delta_{ij}$, which results in $ \mathbf{v}_s = \mathbf{v}_0 $. Even if the elite is not able to make a way through the crowd, the elite can achieve its desired direction and speed through collective drift. It is interesting to find that the effect of local microstructure becomes irrelevant for very large $ K $ and desired motion $ \mathbf{v}_0$ of the elite is forced on every agent in the system. Similarly, when $ K\to 0 $, we recover the motion of a frozen crowd with just the elite agent propelling $ \mathbf{v}_s=\frac{\mathbf{v}_0}{N}$. 

For intermediate values of $ K $, where the effect of microstructure is important, it is interesting to ask how much of this microstructure, around the elite, actually contributes to the drifting direction and magnitude. To answer this question, we analyze the $ r $ and $ \theta $ dependence of the dipole force. Let's begin with the $ r $-dependence. Agents far away experience only a smaller magnitude of force $ \sim \frac{1}{r^2} $. However, the number of agents at a distance $ l $ from the elite has an $ r^2 $ dependence. Hence the total force would be $ r^2\times\frac{1}{r^2}=\mathcal{O}(1) $. This suggests that drifting depends on the microstructure of the entire assembly. Turning to the $\theta$-dependence, for the form of the dipole force used (Eq \ref{dipolefield}), the integral around the elite at a distance $ l $ from it, is zero $ \big(\int_{\theta} \mathbf{U}_{\mathbf{r}_j=l} d\theta=0\big)$. However, because of the granularity of the agent assembly around the elite (since inert agents are comparable in size to the elite), and the absence of symmetry (due to the perturbations given to the initial conditions), the $\theta$ positions around the elite are not well represented. As a result, the integral does not vanish $ (=0) $ giving rise to a non-zero value for sum of dipole forces.  However, as $ l $  increases, agents positions become less granular and the integral may take smaller values. This is sensitively dependent on how the agents are frozen around the elite. Only when the distance is much larger than the length scale in the system $ l\gg R $, the integral might vanish. This large $ l $, would correspond to a system size $ \gg (10\times N) $, much greater than that considered in this work.

To understand the effect of microstructure further, we turn to numerical simulations. Once the agent-assembly is frozen, for high densities, one can compute the $ I_{ij} $ for each agent in the frozen assembly, and estimate the drift velocity $ \mathbf{v}_s $ from Eq \ref{coll-drift-sum-final}. Instead of using all the information of the inert-agent positions, we compute the $ \mathbf{v}_s $ from partial information of the microstructure, with only agents around the elite, say $ r=C_k $. The different distances $ C_k $ considered are $ C_1=3 R $, $C_2 = 6 R $, $ C_3=\frac{1}{2} W $ and $C_k>L$ corresponding to the entire microstructure. In addition, we also consider the effect of the images of the elite for $C_k>L$. Figure \ref{fig:Colldriftall} shows the drift velocity of the assembly observed in simulation and that which is computed from both the partial and complete knowledge of the frozen microstructure. The three configurations presented are exemplary of the drifting frozen assemblies observed in our simulations. We see that $ \mathbf{v}_s $ computed from the microstructure information withing $ C_1 $ seldom predicts the correct direction or the magnitude. 

Clearly as $C_k$ increases, the predictions become better. Deviations from the observed drifting, even for large values of $ C_k $, can be attributed to the departure from the assumptions made to perform these calculations: steady-state and frozen-ness of the assembly. It is interesting to note the large variation in the observed drifting even among the select few examples in figure \ref{fig:Colldriftall}. When comparing figures \ref{fig:Colldriftall} (a) and (b) (regular packed initial configurations), we observe an overall collective drift in the direction that the elite desires to move in. However, we observe that speeds of drifting differ by a factor of $2$. On the other hand, when one begins with a disordered assembly where the heterogeneity in the density of packing and the corresponding variation in the microstructure is high, both the direction and magnitude can be very different (as in figure \ref{fig:Colldriftall} (c)). One should note that it is impossible to predict the direction of drifting from only a visual observation of the assembly of the inert crowd.

The 2D periodicity of the domain allows the agents to drift in any direction. If the assembly was 1D periodic (say in $\mathbf{e}_{\pm x}$, parallel to the desired direction of the elite), bound by walls in the perpendicular direction ($ W\mathbf{e}_{\pm y}$), drifting would have been constrained to $ \mathbf{e}_{\pm x} $ directions. This is because the walls, by definition, are immovable and hence are capable of absorbing forces in the $ \mathbf{e}_y $ direction.

\subsection{Mobility of the elite}

The mobility of the elite agent computed from Eq \ref{Mobility_X} includes both, 
\begin{inlineenum}
	\item the motion of the elite due to re-arrangement of the inert agents $ \mathcal{M}_e $, and
	\item the drifting of the entire assembly when frozen, $ \mathcal{M}_d $, both in the presence and absence of the traffic rule.
\end{inlineenum} 
To compute $ \mathcal{M}_d $, we analyze the variations in the total area of the first Voronoi neighbors of the elite agent. The time-stamps $ t_d $ corresponding to drifting of the assembly are identified (corresponding to negligible or no variations in the Voronoi areas) and the quantity $ \langle \mathbf{v}_e \cdot \mathbf{e}_x \rangle\Big|_{t \in t_d} $ is evaluated at $ t_d $, to determine the mobility $ \mathcal{M}_d $. 

\begin{figure}
	\centering
	\includegraphics[width=1\linewidth]{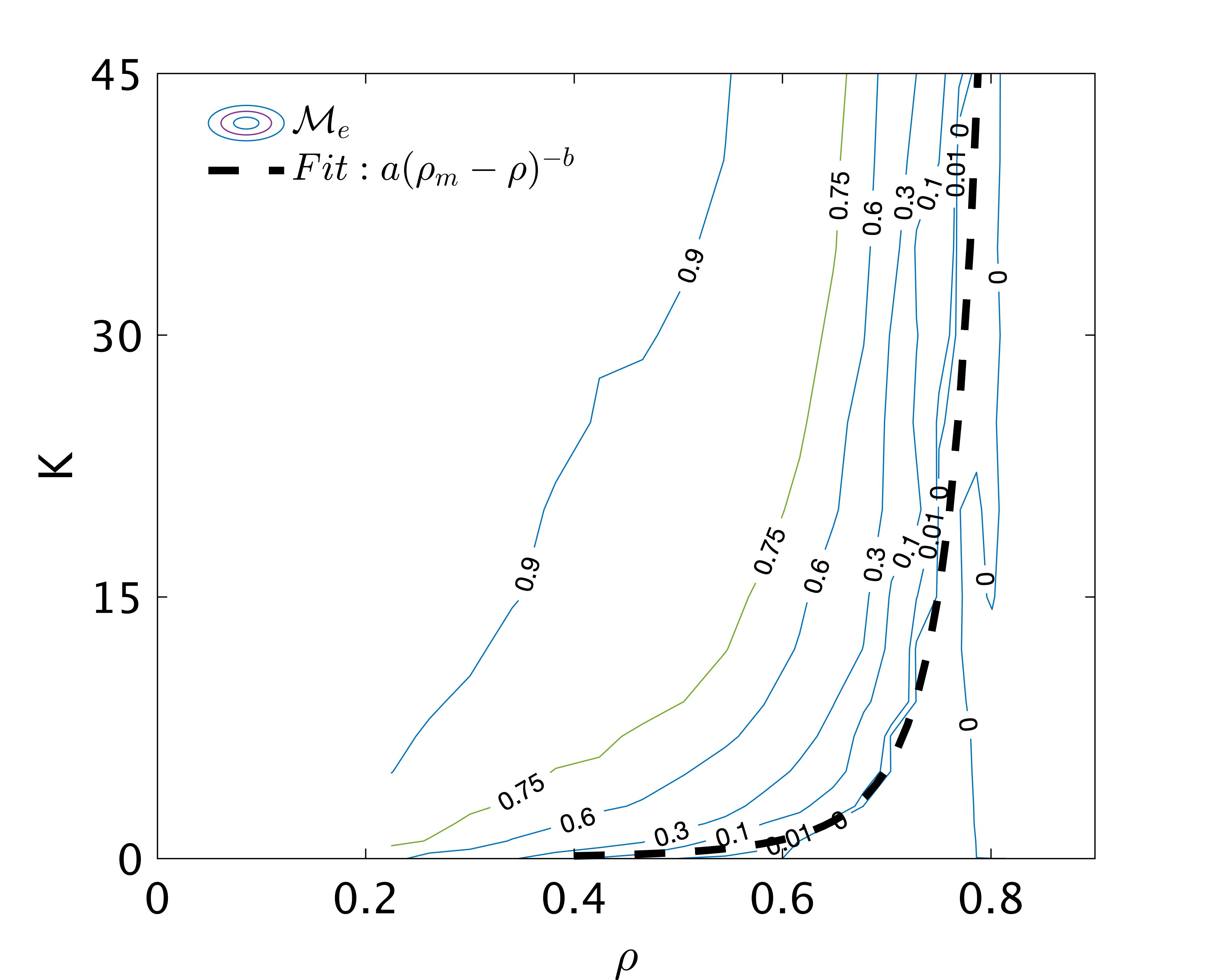}
	\caption{Contours of the mobility of the elite agent, $\mathcal{M}_e$ without any contributions from drifting. The dotted line corresponds to the fit ($\alpha = 0.01$ and $\beta = 4$) as in Eq \ref{MobilityContourFits}.}
	\label{fig:mobilityonlyelitewithfit}
\end{figure}

Previously in figure \ref{fig:paramaterspaceregularpacking}, where the four distinct regions are marked, we show two sets of contours: the thick lines belong to $ \mathcal{M} $, the total mobility and the thin ones are $ \mathcal{M}_d $ that are only due to collective drifting. Any mobility of the elite in region IV is only due to $ \mathcal{M}_d $, whereas in high-K part of region II, mobility is due to both $ \mathcal{M}_d $ and $ \mathcal{M}_e $.
Hence, one can subtract the contribution of $ \mathcal{M}_d $ from the total mobility $ \mathcal{M} $, to compute $ \mathcal{M}_e $. Figure \ref{fig:mobilityonlyelitewithfit}, shows the contours of $\mathcal{M}_e$.

At a certain crowd-density $ \rho_0 $, in the absence of the traffic rule $ K=0 $, the elite agent would have a corresponding mobility $ \mathcal{M}_e =  \mathcal{M}_0$. If the density of the crowd were to increase, so will the  dipole strength $ K $ required to maintain the same mobility $\mathcal{M}_0$. However, the crowd cannot be packed to a density greater than the maximum $ \rho_m=\frac{\pi}{2\sqrt{(3)}} $.,which also means that close to this value of packing density no amount of dipole force can facilitate motion of the elite.  Hence, we expect $ K$ to diverge at $\rho_m$, i.e., $ \rho \to \rho_m $, $ K_c\to \infty $. This can be captured nicely by the functional form relating $ K_c \text{ to } \rho$ as shown in Eq \ref{MobilityContourFits}. 
\begin{equation}\label{MobilityContourFits}
    K_c =\frac{\alpha}{(\rho_m - \rho)^{\beta}}
\end{equation}
For very small values of $ \mathcal{M}_0 \to 0 $, the mobility curves start to get dense and can be sufficiently explained by Eq \ref{MobilityContourFits} with $\beta$ as high as $ \simeq 4 $ and $\alpha$ is $ \mathcal{O}(10^{-2}) $ (the fit is shown as a thick dotted line in figure \ref{fig:mobilityonlyelitewithfit}). 

\begin{figure*}
    \centering
    \includegraphics[width=0.95\linewidth]{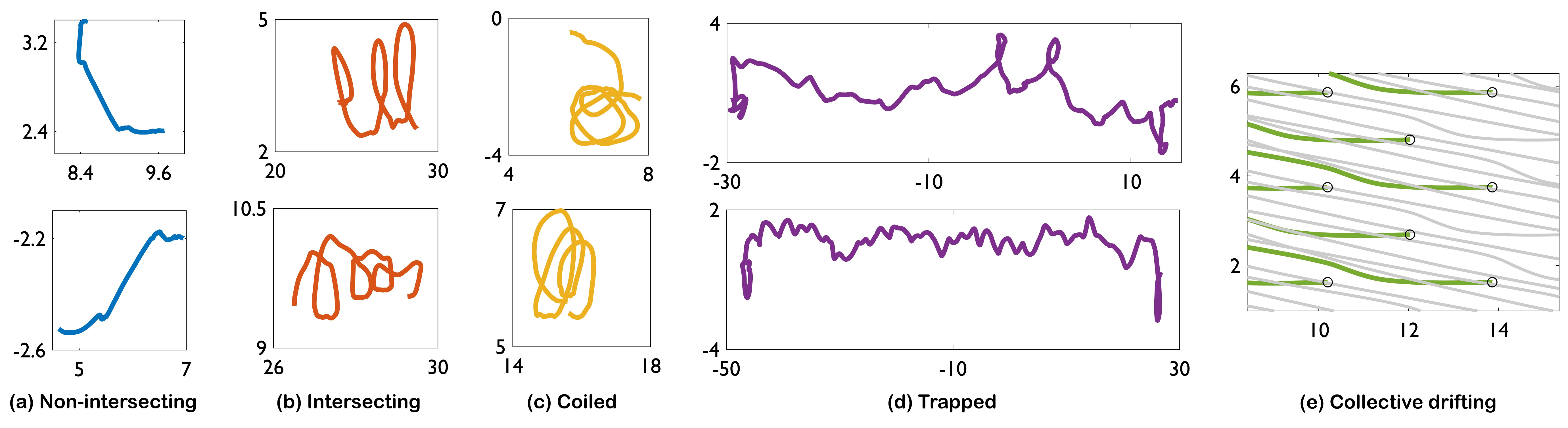}
    \caption{Representative trajectories of the inert agents in the crowd.}
    \label{fig:trajectories-inert}
\end{figure*}

\subsection{Trajectories of the inert}
While the primary objective of the traffic rule is to maximise the movement of the elite agent through the crowd, the secondary objective is to achieve the primary at minimum movement of the inert. Hence, it is important to understand how the inert move around the elite to achieve the elite's mobility. We explore the trajectories formed by the inert agent as the elite moves past them and classify them broadly into five different types.

\paragraph{Non-intersecting:}
When there is no traffic rule in place, the elite pushes the inert away as it tries to move through the crowd (at low packing densities). This results in trajectories as shown in figure \ref{fig:trajectories-inert}a.

\paragraph{Intersecting:}
In the presence of a traffic rule, inert agents respond to the elite agent differently depending on their relative position with respect to elite. An agent in front of the elite is pushed away, while that in its rear position is pulled towards it. Hence, as an elite agent passes by, an inert moves away and then towards the elite, giving rise to trajectories that intersect, like the one seen in figure \ref{fig:trajectories-inert}b. Multiple crossings observed in the figure is due to multiple encounters with the elite (as a result of the 2D periodic domain).

\paragraph{Coiled:} When the inert agents are slightly away from the elite in the transverse direction, they exhibit coiled trajectories as shown in figure \ref{fig:trajectories-inert}c which results in very minimal displacement to make space for the elite to pass by.

\paragraph{Trapped agents:}
In a small range of densities $ \sim (0.42, 0.55) $ and $ K > 20$ we observe trapping of inert agents in the \textit{sink} part of the dipole field of the elite. When densities are lower, the effect of the traffic rule is also low because the elite's velocity is close to $\mathbf{v}_0$ (see Eq \ref{dipolefield}). However, as the density increases, $ \mathbf{v}_e $ reduces and consequently, the traffic rule strength becomes higher and affects the motion of agents in the neighborhood significantly---with the inert attracted to the sink and repelled from the source. Inert agents get trapped in the rear end of the elite and follow it as it moves through the crowd. As the density is increased further, the inert experiences more collisions with other inert agents in the crowd which dislodges the inert from the elite's tail. This phenomenon leads to the formation of long trajectories like the ones shown in figure \ref{fig:trajectories-inert}d. From the perspective of the secondary objective of the traffic problem, trapping should be avoided.

\paragraph{Collective drifting:}
When the system as a whole exhibits drifting due to the traffic rule (section \ref{sec:colldrift}), we observe all agents to move in a similar manner. Figure \ref{fig:trajectories-inert}e, shows how all the inert agents have similar trajectories during drifting. Slight differences could be due to minor re-arrangements that occur while the frozen assembly drifts. Collective drifting fails to fulfil the secondary objective of minimising the movement of the inert. But one should note that in the absence of any drifting, the assembly is frozen and does not fulfil the primary objective of the problem which is to facilitate the motion of the elite.

\section{Conclusions}
As we slowly enter the age of autonomous vehicles, there is a need to deal with the challenges associated with decentralized-control. Control algorithms in vehicles should not just be designed to avoid collisions, but also react to a time of emergency, like the one described in the introduction of the article---ambulance rushing to the hospital---where there is a need to facilitate motion of the vehicle with priority. 
Identifying the optimal control strategy for the above problem or any other traffic problem for that matter, can be a very hard problem to solve, since we are dealing with many agents.
Hence, swarm engineering has always reached out to bio-inspired methods for answers. The work presented in this article, is an example of a nature-inspired method to the solution of a complex-traffic problem. The traffic rule presented in this article circumvents the use of an optimization approach to scheduling the movement of the inert agents around the elite. However, it implicitly encodes two important objectives that are relevant for the traffic problem: increasing the movement of the elite agent with only minimal movement of the inert agents.

\paragraph*{Summary:} We study the problem of movement of an elite agent with priority through an otherwise inert crowd. Taking advantage of the similarities in granular and traffic flows, we propose a solution from the knowledge of the flow fields of grains around an active intruder. We derive a traffic rule that the inert agents can follow and make small movements to allow the passage of the elite agent. The inert agents respond to an imaginary dipole field around the elite and rearrange accordingly to create enough space for the elite to move. We find that depending on the packing density of the crowd and the strength of the dipole field, the parameter space can be divided into different regions with characteristic flows. We show that the motion of the elite with the traffic rule is more efficient than without it. As density increases, the motion of the elite is via the local melting of the crystallized assembly. At even higher densities, where melting is not observed, motion can be due to a collective drifting mechanism, which depends on the local configuration of the inert around the elite.

\paragraph*{Can the rule be used in real Indian traffic scenario?:}
The traffic rule facilitates the inert agents to move around the elite creating space in front of it and occupy the space already formed at the elite's wake as it propels forward. 
It is interesting to note that in real lane-less flows, when an ambulance tries to make its way through a crowd, vehicles in the front of it begin to form denser arrangements, while those at the back readily occupy the empty space in the ambulance's wake, without giving any opportunity for the transport of the available space from the back to the front of the elite agent---much like the case with no traffic rule in our simulations. 
This is because, in real vehicular traffic, the maneuver that the inert have to follow according to the traffic rule discussed in this article, is close to impossible to achieve for human drivers under normal conditions. 
However, it could be a possibility when the entire crowd is in motion, where the inert can simply slow down and drift sideways to occupy the space behind the elite. When the agents are autonomous and self-driving, it may be possible to implement this traffic rule.

\paragraph*{Other problems in swarm robotics:}
The above idea can be extended to other interesting problems in robotics, namely the movement of a drone or a swarm of them, through an array of obstacles. The problem here, is to figure out the optimal flight path that results in efficient movement of the drones. If a dipole field is assumed to be centered around a perceived obstacle, it will help in guiding the motion of drones around it. In the case of multiple obstacles, a simple superposition of all the fields can essentially provide a possible safe path for the drones. It would be interesting to see how this rule would fare, in a virtual environment in the presence of collision avoidance measures and then eventually test these nature-inspired rules on real swarm robots.

\begin{acknowledgments}
DRM would like to thank DST INSPIRE faculty award for the funding.
\end{acknowledgments}

\bibliography{library.bib}

\end{document}